\documentclass[english,11pt,aps,prd,letterpaper,preprintnumbers,floatfix,nofootinbib,showpacs,superscriptaddress, notitlepage]{revtex4-1} \usepackage{amssymb}
\usepackage{slashed}
\usepackage{amsmath}
\usepackage{mathtools}
\usepackage{float}
\usepackage[utf8]{inputenc}
\usepackage{graphicx}
\usepackage{latexsym}
\usepackage{rotating}
\usepackage{epsfig}
\usepackage{xcolor}
\usepackage{natbib}
\usepackage[colorlinks=true, pdfborder={0 0 0}]{hyperref}
\usepackage[normalem]{ulem}

\definecolor{linkcolor}{rgb}{0,0,0.5}

\newcommand{\psicp}{\ensuremath{\psi_\mathrm{CP}}}
\newcommand{\Dphi}{\ensuremath{\Delta \phi}}

\begin{document}

\title{Measuring the CP properties of the Higgs sector at electron-positron colliders}

\author{I. Bo\v{z}ovi\'{c}-Jelisav\u{c}i\'{c}}{\thanks{ibozovic@vin.bg.ac.rs}}
\affiliation{\say{VIN\u{C}A} Institute of Nuclear Sciences, Belgrade}

\author{N. Vuka\u{s}inovi\'{c}}
\affiliation{\say{VIN\u{C}A} Institute of Nuclear Sciences, Belgrade}

\author{D. Jeans}
\affiliation{Institute of Particle and Nuclear Studies, KEK}
\affiliation{The Graduate University for Advanced Studies, SOKENDAI}

 \begin{abstract}

The violation of CP symmetry at the electro-weak scale is one of the essential ingredients for electro-weak baryogenesis.
It is therefore of great interest to map the CP properties of the Higgs sector in as much detail as possible.
A linear electron-positron Higgs factory collider will provide many opportunities to probe the CP nature of the Higgs sector,
thanks to access to several Higgs production processes at a wide range of centre-of-mass energies.

In this paper we report on two studies based on full simulation of detector response and realistic reconstruction algorithms: 
1) a study of $h \to \tau \tau$ at ILC-250, in which mixing of CP eigenstates can be measured to a precision of 75~mrad; and 
2) the current status of an ongoing study of the $ZZ$--fusion process at 1.4~TeV CLIC and 1~TeV ILC, for which we expect to achieve concrete results during the Snowmass study period.


\end{abstract}

\maketitle

\vskip -0.3in 
\begin{center}
\rule[-0.2in]{\hsize}{0.01in}\\
 \vskip 0.1in 
 Submitted to the  Proceedings of the US Community Study\\
 on the Future of Particle Physics (Snowmass 2021)\\
 \rule{\hsize}{0.01in}\\
\end{center}


\def\thefootnote{\fnsymbol{footnote}}
\setcounter{footnote}{0}

\newpage

\section{Introduction}

The violation of CP symmetry at the electro-weak scale is one of the essential ingredients for electro-weak baryogenesis.
It is therefore of great interest to map the CP properties of the Higgs sector in as much detail as possible.

CP violation in the Higgs sector can be probed in both the fermionic and bosonic couplings of the Higgs boson.
To cover the various measurements, it is highly advantageous to utilize collisions at a wide range of centre-of-mass energies. 
Linear electron-positron ``Higgs factory'' colliders, with their inherent extendibility in energy, provide
the ability to carry out a complete survey of the sector.

In this paper we discuss the measurement of CP effects in $h \to \tau\tau$ in 250~GeV collisions at ILC~\cite{ILC}, where the analysis
profits from the experimental advantages of the Higgs-strahlung process~\cite{khoze}.
We also discuss the $ZZ$-fusion production process, which provides a powerful tool to probe the bosonic coupling.
The $ZZ$-fusion cross-section is tiny at 250~GeV but grows with energy, so a high energy collider is imperative. 
We report on the status of ongoing studies at 1.4~TeV CLIC and 1~TeV ILC.

High energy collisions above $\sim550$~GeV also allow measurement of the top-quark's Yukawa coupling in the process
$e^+ e^- \to t \overline{t} h$; we leave a detailed study of this channel's CP sensitivity as a future exercise for the interested reader.

\section{International Large Detector and CLIC detector}

The studies presented here are based on detailed Geant-4 simulations of ILC's International Large Detector concept (ILD)~\cite{ILD} and the CLIC-ILD concept~\cite{Linssen:2012hp}.
To meet the strict requirements to maximize the physics output at these facilities, these detectors are designed
with an ultra-precise and light-weight vertex detector, 
a high-precision low mass tracking system based on a TPC complemented with silicon layers
operating in a 3.5-4~T magnetic field, and calorimeters with highly segmented readout to enable optimal 
particle flow reconstruction. Almost the entire $4\pi$ solid angle around the collision point is covered
by active detectors. Such detectors can provide impact parameter resolution of $\sim3~\mu m$ and resulting excellent
jet flavor tagging, track momentum resolution at 
high momentum of $\delta p_T / p_T \sim 3 \times 10^{-5} p_T$,
and jet energy resolution of $3-4\%$ over a wide range of jet energies.

Signal and background processes are generated using the WHIZARD event generator~\cite{whizard, omega} taking into account
the accelerators' beam parameters, including such effects as beamstrahlung and initial state radiation.
Generated events are then passed through detailed Geant4~\cite{g4} models of the detectors, described using 
tools provided by DD4hep~\cite{dd4hep}. The energy deposits calculated in the Geant4 simulation
are digitised and reconstructed using the tools provided in the ilcsoft software framework~\cite{ilcsoft},
including particle flow reconstruction by PandoraPFA~\cite{pandora}.

\section{CP measurement in $h \tau \tau$ at ILC-250}

CP violation in the tau lepton's Yukawa coupling can explicitly be introduced by a Lagrangian term such as 
\begin{equation}
\mathcal{L}_\mathrm{H\tau\tau} = g \overline{\tau} ( \cos \psicp + i \gamma_5 \sin \psicp ) \tau \mathrm{H}
\label{eqn:cpv}
\end{equation}
which is CP-conserving
for $\psicp = 0$ and maximally violates CP for $\psicp = \pi/2$. 
The CP nature of the tau-pair produced in Higgs decay is reflected in the correlation between 
the transverse tau spin components, and thereby in the relative distribution of the taus' decay products.
This analysis demonstrates how to reconstruct the tau spin orientation from its decay products,
and to extract the relative CP even and odd contributions in a sample of $h \rightarrow \tau \tau$ events.

\begin{figure}
\centering
\includegraphics[width=0.49\textwidth]{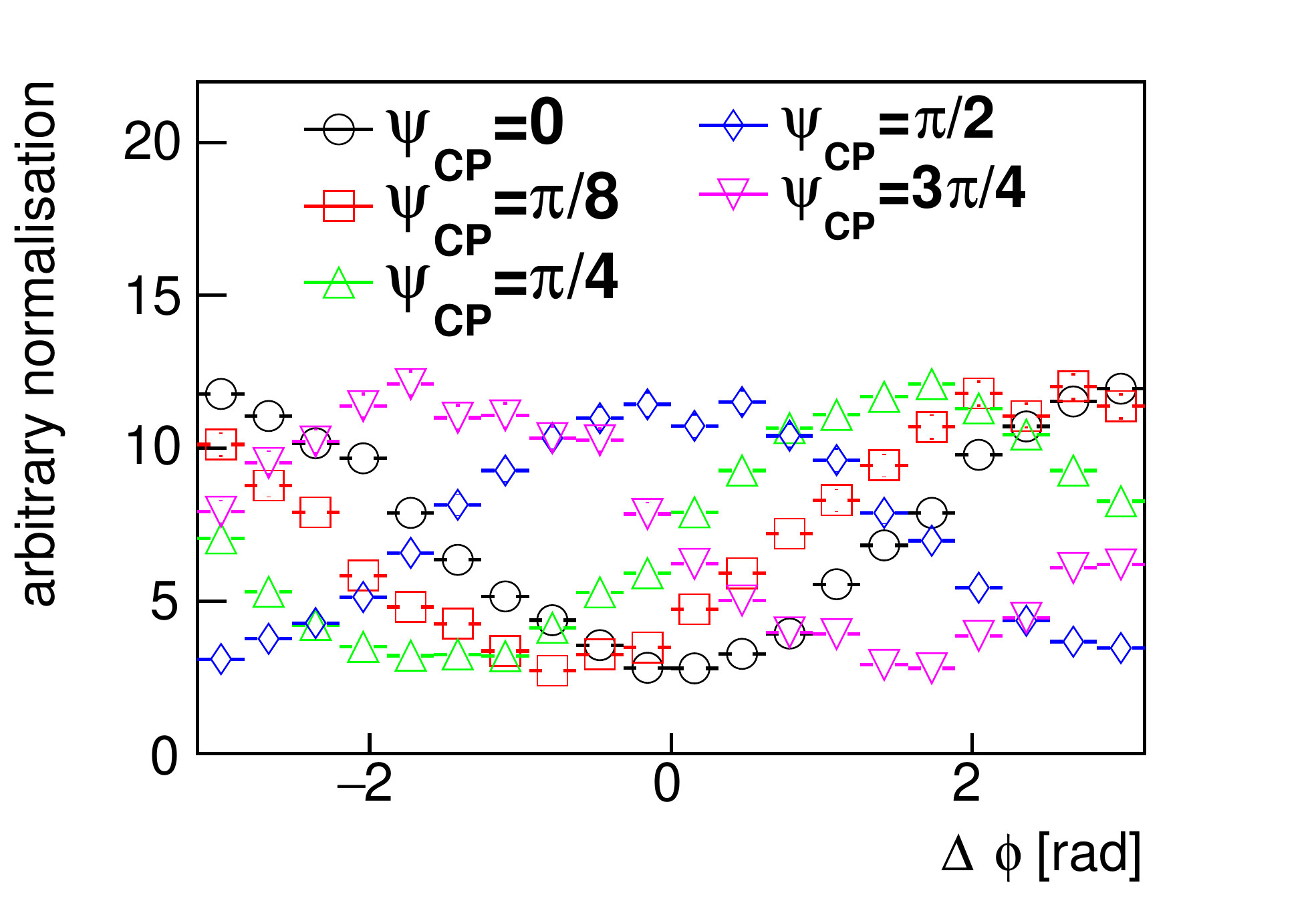} 
\caption{
Distributions of events in $\Dphi$, integrated over $\theta^\pm$, at MC truth level, for different values of \psicp~\cite{taucp}.
}
\label{fig:mc_dphi}
\end{figure}

An optimal tau polarimeter, or spin orientation estimator, is easy to define in decays to
one or two pions, which together account for around 36\% of tau decays.
Considering the Higgs rest frame, the CP sensitive observable \Dphi\ is defined as the difference in azimuthal angle, 
defined with respect to the tau momentum direction, between the two polarimeters.
The distribution of events in \Dphi\ is sinusoidal, with a phase which is sensitive to the 
admixture of even and odd CP components. Examples of distributions of this CP-sensitive observable at
different CP mixing angles \psicp\ are shown in Fig.\ref{fig:mc_dphi}.

\subsection{Analysis in ILD}

We here summarize the analysis described in~\cite{taucp}.
The signal process is $e^+ e^- \rightarrow Z H$, with subsequent decay of the Z boson to electrons, muons,
or hadrons, and of the Higgs boson to a pair of tau leptons. 
The signal tau leptons are decayed by Pythia8~\cite{pythia8}
into the one- and two-pion decay channels, taking spin correlations into account.
Considered backgrounds include two- and four-fermion final states, and well as other 
decay modes of the Higgs-strahlung process.

\subsection{Tau lepton reconstruction}

We here outline the tau reconstruction procedure described in~\cite{taureco}.
Each tau lepton in the final state decays to at least one neutrino which escapes detection.
To optimally reconstruct polarimeters and CP mixing, this neutrino momentum should be
reconstructed. Assuming that both taus decay hadronically, there are 6 unmeasured quantities per
event, corresponding to the 3-momenta of the event's two neutrinos. 
By considering the measured momenta of visible tau decay products, their impact
parameters with respect to the reconstructed tau production vertex, and imposing 
constraints on the reconstructed tau mass and the transverse momentum of
the di-tau system, the neutrino momenta can be reconstructed up to a 4-fold ambiguity,
which can be resolved in many cases by considering the reconstructed tau lifetime.

\subsection{Event selection}

Kinematic observables, for example the di-tau invariant mass, the mass recoiling against
the Z decay products, and the angle between the total event momentum and the beam axis, were used 
to select an event sample enriched in signal events. Three different channels were
considered, according to the Z boson decay into electrons, muons, or hadrons. 
Assuming a total integrated luminosity of 2/ab at ILC-250, a few tens of events are
expected in the case of Z decays to electrons and muons, with a signal-to-background ration of around 1:5.
For hadronic Z decays, over 500 signal events are expected after selection, but with much larger
background contamination.

The tau decay mode is reconstructed by considering the number of reconstructed charged hadrons and
photons within the candidate tau jet, as well as the invariant mass of various groupings of those
particles. The efficiency to correctly identify both tau decay modes in a selected event ranges between
87\% and 94\%, depending on both the Z and tau decay modes.

\subsection{CP sensitivity}

\begin{figure*}
\centering
\includegraphics[width=0.49\textwidth]{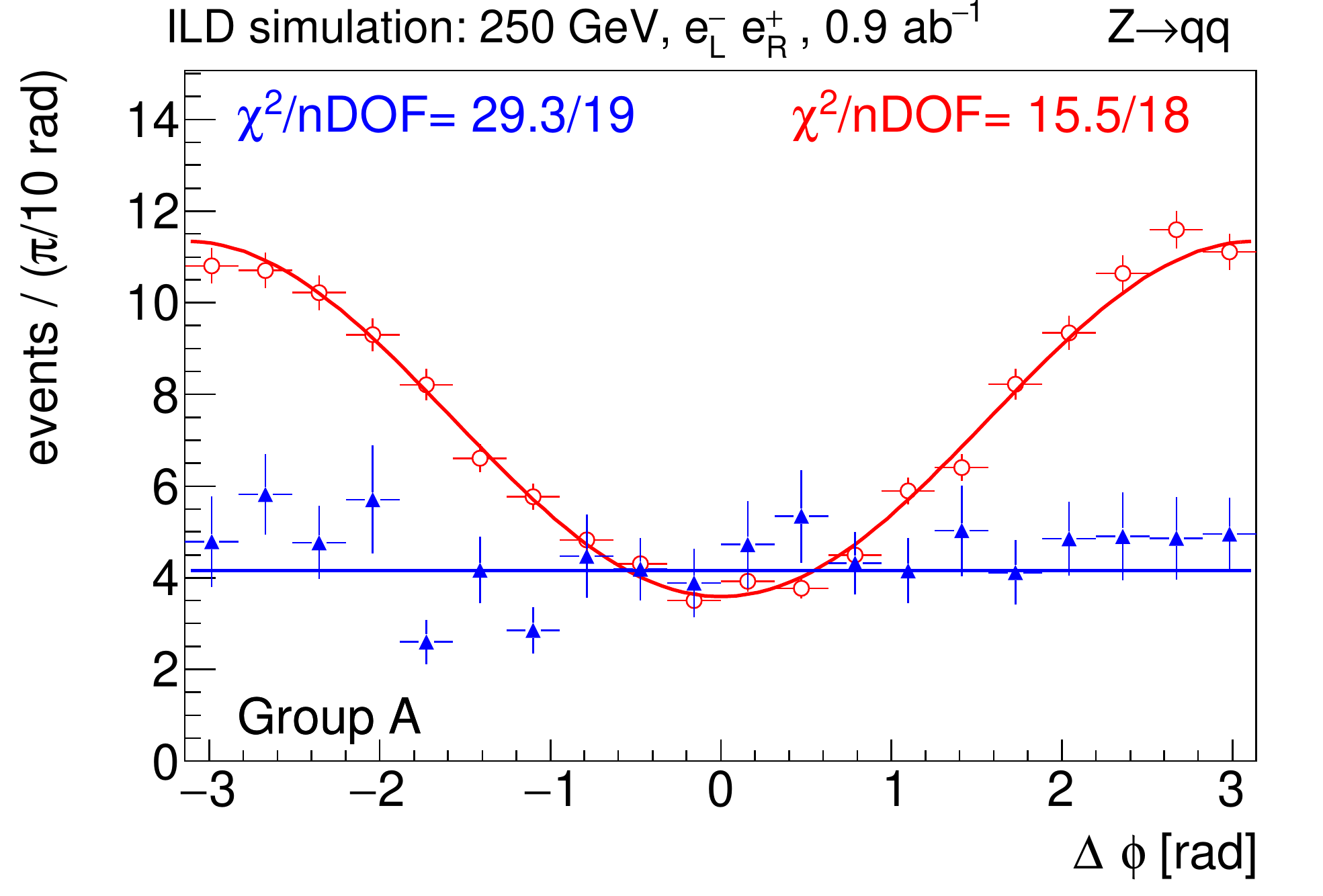}
\includegraphics[width=0.49\textwidth]{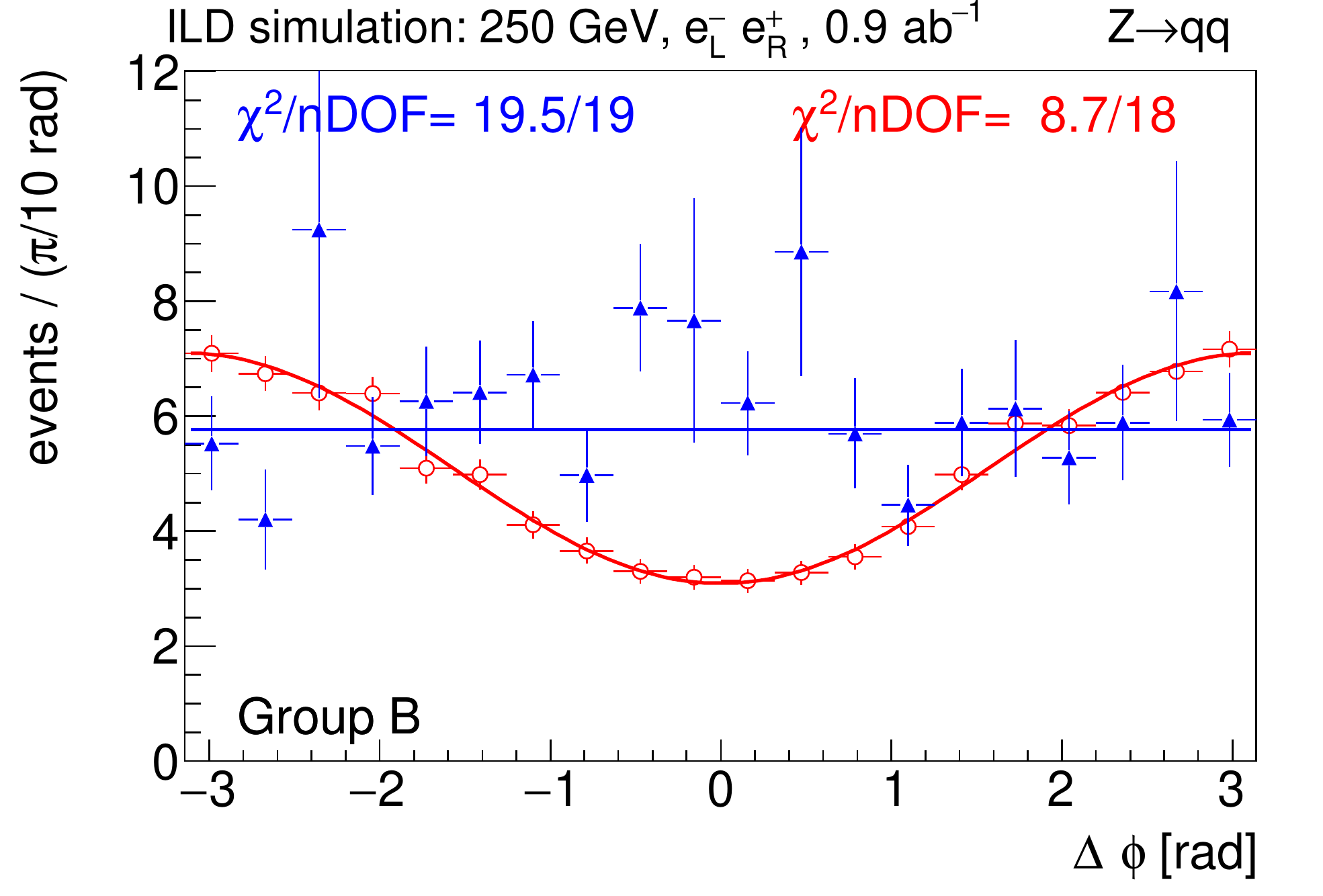}
\caption{Distributions of reconstructed \Dphi\ in the ``gold'' (group A) and ``silver'' (group B) categories,
for events selected in the hadronic Z decay channel. 
The distribution of signal events is shown as open red circles, and the 
total background distribution as filled blue triangles.
Signal samples were generated with $\psicp=0$ (i.e. the SM)~\cite{taucp}.
}
\label{fig:dphi}
\end{figure*}

At this stage the tau polarimeters can be calculated from the tau decay products' momenta boosted
into the tau rest frames. Events are classified according to their expected sensitivity to 
CP effects, depending on the orientation of the polarimeter with respect to the
tau momentum, the quality of the impact parameter reconstruction, encoded by its significance, 
and the expected contamination both from background processes and mis-identified tau decay modes.
Four classes are defined, which in this Olympic year we label as ``gold'', ``silver'', ``bronze'' and ``the rest'',
in each of which the CP-sensitive observable \Dphi\ is fitted using a maximum likelihood fit 
to extract the CP mixing parameter, as shown in Fig.~\ref{fig:dphi}.

In the case of 1/ab of data with unpolarized beams, the expected sensitivity on the 
CP mixing angle $\delta \psicp$ is 116~mrad, dominated by events with hadronic Z decay and with at least
one tau decay to two pions. If we assume that all background can be rejected, the sensitivity
improves to 76~mrad, while if in addition all signal events are selected and polarimeters are
perfectly reconstructed, the ultimate sensitivity is 25~mrad.

Considering the 2/ab expected at ILC in various beam polarisation combinations, the
expected sensitivity on $\delta \psicp$ is 75~mrad when all experimental effects are taken into account.
There is considerable potential scope for improvement, for example by improved reconstruction
and selection methods, better tau decay mode identification, and the use of additional tau decay modes.

\section{CP measurement of the $h Z Z$ vertex}

We here report on the current status of an analysis of the CP properties of the $h Z Z$ vertex, using the $ZZ$-fusion
production in high energy electron-positron collisions at CLIC (1.4~TeV) and ILC (1~TeV).

\subsection{CPV mixing in the Higgs sector}

Let us assume that SM-like Higgs boson is a quantum superposition of CP-even (scalar) and CP-odd (pseudoscalar) states through CP violating mixing angle $\Psi_\mathrm{CP}$. 
The Higgs to $Z$ boson tensor coupling can be then described as:

\begin{equation}
\label{eq1}
g_{HZZ} = ig\frac{ M_{Z} }{ cos\theta_{W}} (\lambda_{H} \cdot g^{\mu\nu} + \lambda_{A} \cdot \epsilon^{\mu\nu\rho\sigma}  \frac{(p_1 + p_2)\rho(p_1 - p_2)\sigma}{M^2_{Z}}) 
\end{equation}

\noindent where $\mathrm{p_1}$ and $\mathrm{p_2}$ are four-momenta of the $Z$ bosons \cite{r2}. Contribution to gauge $HZZ$ coupling
with $\mathrm{\lambda_H}$ has a structure of the CP-even SM Higgs boson coupling (H), while the term with $\mathrm{\lambda_A}$
stands for the CP-odd, pseudoscalar coupling. Superposition of the two states can be described as:

\begin{equation}
\label{eq2}
h = \lambda_H \cdot H + \lambda_A \cdot A
\end{equation}

where
\begin{equation}
\label{eq3}
\lambda_H = \cos\psi_{CP} \hspace{1cm} \mathrm{and} \hspace{1cm} \lambda_A = \sin\psi_{CP}
\end{equation}

Although we further consider Higgs to $b\bar{b}$ decays in order to minimize the list of possible backgrounds for this particular event topology, 
we do not make specific assumptions on Higgs couplings to fermions, thereby preserving the model independency.

\subsection{Sensitive observables}

\begin{figure}[h]
\centering
\includegraphics[width=.4\textwidth]{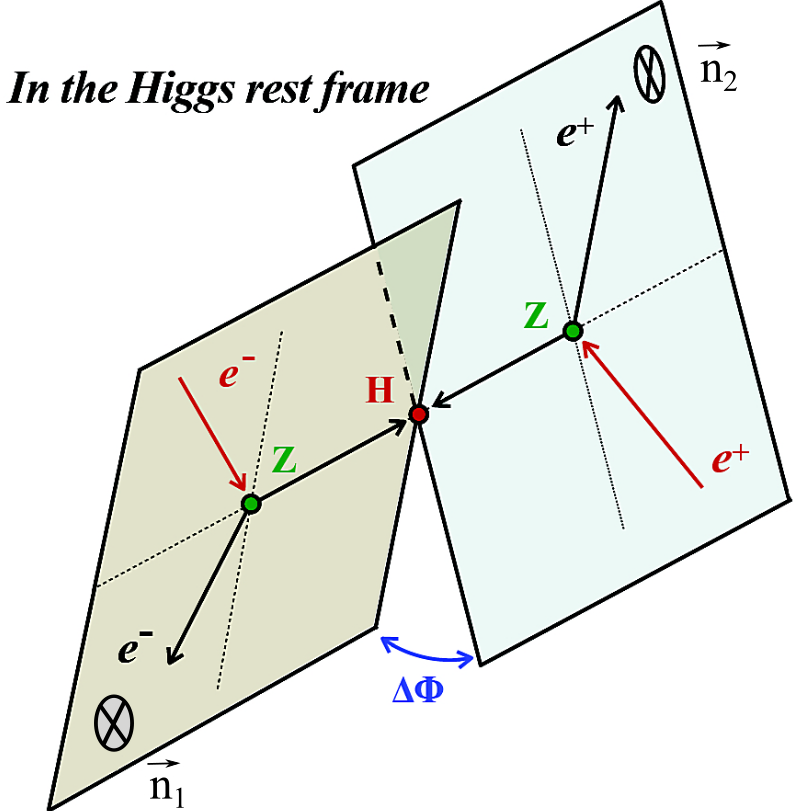}
\caption{Illustration of the CP sensitive angle $\Delta\Phi$ in the $ZZ$ fusion process.}
\label{fig1}
\end{figure}

We are considering Higgs boson production in $ZZ$-fusion $e^+e^-\rightarrow e^+e^-H (H \rightarrow b \bar{b})$,
with the 2-jet and two electron spectators experimental signature. According to \cite{r3}, there are
several observables sensitive to the mixing from Eq. \ref{eq3}. We further use the angle $\mathrm{\Delta\Phi}$ between
production planes (Fig. \ref{fig1} and Eq. \ref{eq4}).

\begin{equation}
\label{eq4}
\Delta\Phi = \arccos(\hat{n}_{1}\cdot\hat{n}_{2})
\end{equation}

where

\begin{equation}
\label{eq5}
a = \frac{ q_{Z_{e^-}}\cdot (\hat{n}_{1} \times \hat{n}_{2})}{ |q_{Z_{e^-}}\cdot (\hat{n}_{1} \times \hat{n}_{2})|}, \hspace{1cm} \hat{n}_{1} = \frac{ q_{e^-_{i}}\times q_{e^-_{f}} }{ |q_{e^-_{i}}\times q_{e^-_{f}}| } \hspace{1cm} \mathrm{and} \hspace{1cm} \hat{n}_{2} = \frac{ q_{e^+_{i}}\times q_{e^+_{f}} }{ |q_{e^+_{i}}\times q_{e^+_{f}}|}
\end{equation}

\noindent and q stands for the corresponding momenta, while $\mathrm{\hat{n}_{1,2}}$ define unit vectors orthogonal to the
production planes. Parameter $a$ defines how the second (positron) plane is rotated w.r.t. the
first (electron) plane. If it falls backwards (as illustrated in Fig. \ref{fig1}) $a = -1 $ otherwise $a = 1$.
Direction of Z boson in the electron plane regulates the notion of direction (forward or backward) by the right hand rule.


\subsection{Event samples}

Signal event samples of approximately 8000 events, corresponding to a nominal integrated luminosity of 2.5 $\mathrm{ab^{-1}}$, are reconstructed at 1.4\,TeV CLIC, while about 1000 events
at 1\,TeV ILC are considered, corresponding to only 1/3 of the expected signal statistics in 1\,$\mathrm{ab^{-1}}$ of data. The full detector simulation, including realistic luminosity spectrum 
and overlay of beamstrahlung background is applied in both experiments.

Several background processes with similar final state as the signal are considered, where the major contribution comes from backgrounds with large cross-sections, 
such as $e^+e^-\rightarrow q\bar{q}$ and $e^+e^-\rightarrow e\nu qq$.

\subsection{Method}

\begin{figure}[h]
\centering
\includegraphics[width=.4\textwidth, height=.3\textwidth]{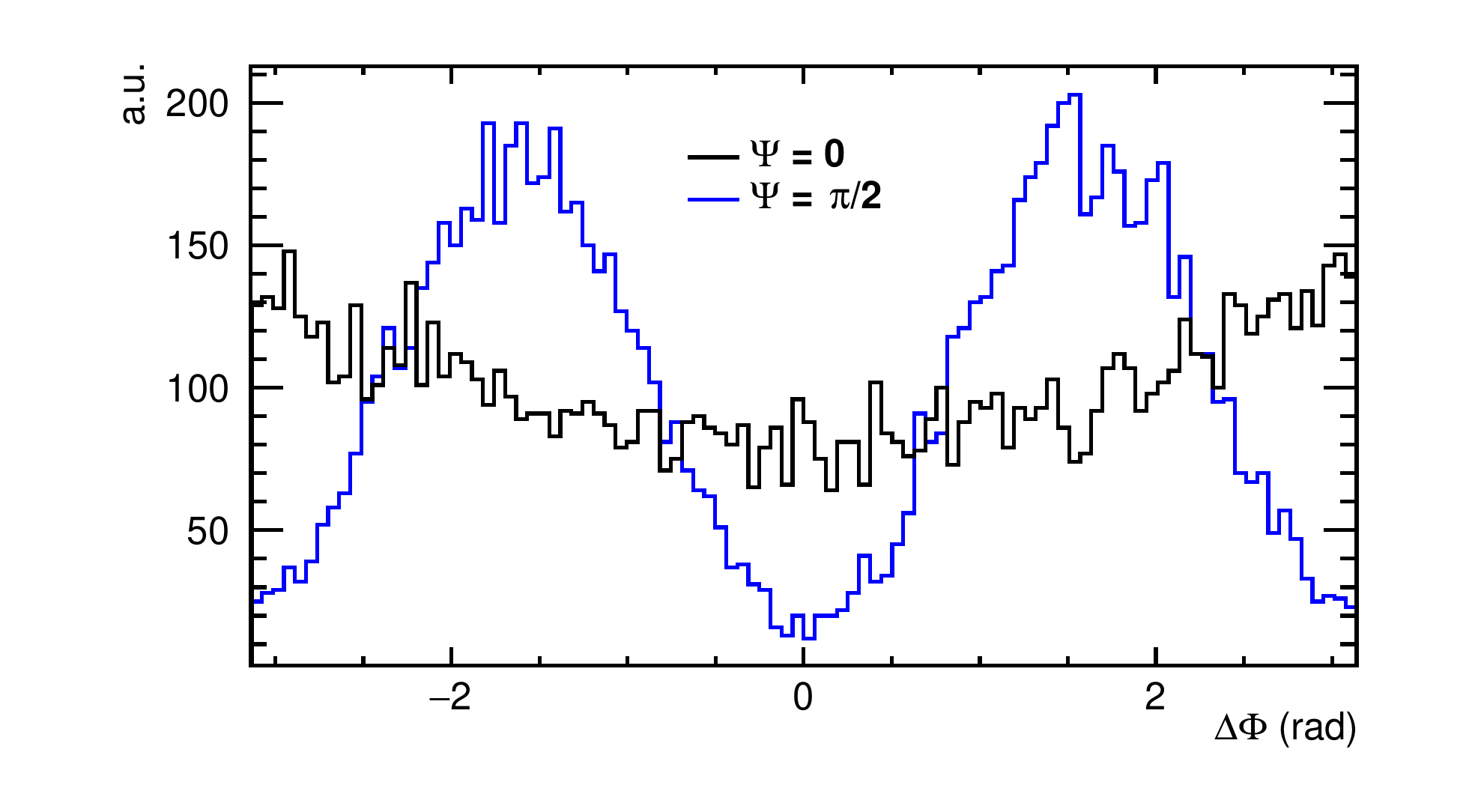}
\caption{ $\mathrm{\Delta\Phi}$ distributions for $\mathrm{\Psi_{CP} = 0}$ and $\mathrm{\Psi_{CP} = \pi/2}$ corresponding to pure scalar and pseudoscalar states of the Higgs boson.}
\label{fig2}
\end{figure}

The method of both analyses is the same:
\begin{enumerate}
    \item It starts with the isolation of electron spectators in the tracking volume, requiring two isolated leptons per event. The simulation, reconstruction and analyses are carried out with the ILCDIRAC framework \cite{r4}. Electron isolation is optimized according to track energy, the ratio of energies deposited in the electromagnetic and hadronic calorimeters, as well as the impact parameters of the electron track. Electron tracks are required to satisfy two-dimensional requirements on cone energy vs. electron energy, where the cone energy sums up all particle energies excluding the electron itself, in a cone of optimized size of few degrees around the isolated lepton track. This is achieved by employing the Isolated Lepton Finder Marlin processor \cite{r5}.
    \item In order to separate signal from background events, multivariate analyses (MVA) based on TMVA toolkit \cite{r6} is employed. 
Several sensitive observables are used as input to MVA, like reconstructed Higgs mass and energies of the final electrons. 
Due to specific properties of the 2e 2-jet final state, considered background is highly suppressed in both ILC and CLIC study, up to a percent level compared to the signal in the CLIC case. 
Although limited in statistics compared to the inclusive Higgs production in $ZZ$-fusion, this approach enables: 
a) combination with other Higgs decay channels (i.e. $H \rightarrow WW$) to improve statistical precision, 
b) almost background free samples simplifying the method in terms of a background fit, at least at CLIC.
    \item Upon obtaining reconstructed $\mathrm{\Delta\Phi}$ distributions, $\mathrm{\Psi_{CP}}$ is extracted from the fit of a single (pseudo) experiment. 
All simulations are done for $\mathrm{\Psi_{CP} = 0}$ that is assuming pure state of the SM Higgs boson H. 
The fit correctly reproduce generated distributions for $\mathrm{\Psi_{CP} = 0}$ and $\mathrm{\Psi_{CP} = \pi/2}$ obtained using 2HDM model in 
WHIZARD V2.8.1 \cite{r7} within the UFO framework, as illustrated in Fig. \ref{fig2}.
\end{enumerate}

\subsection{Status of the ongoing analyses}
At 1.4\,TeV CLIC we are finalizing the step 4) from the previous section, with a promising uncertainty of individual experiments of order of a few degrees for absolute statistical uncertainty of $\mathrm{\Psi_{CP}}$. At ILC, we have completed the step 2), with background
suppression for a factor 10$^{3}$ and CP insensitive behavior.

\section{Summary}

Linear electron-positron colliders will enable a number of measurements sensitive to the CP nature of the Higgs sector,
which is a crucial ingredient in the quest to understand the possibility of baryogenesis at the electro-weak scale.
The clean experimental environment of an electron-positron collider is conducive to precision measurements of the often complex final states,
while the energy-extendibility of a linear collider will allow a number of complementary measurements at different
centre-of-mass energies.
In this paper we have highlighted a Higgs CP measurement using the $\tau$ Yukawa coupling at 250~GeV, where mixing
between CP even and odd contributions can be measured with a precisio of 75~mrad at ILC-250.

Higher energy stages, at a TeV or higher, can be used to measure CP effects in the $HZZ$ coupling by studying the $ZZ$-fusion Higgs production process.
We expect to provide quantitative results on this analysis during the remainder of the Snowmass study.
The combination of such a suite of measurements will allow a global and complete view of the Higgs' CP nature to be established.

\section*{Acknowledgments}

We would like to thank the LCC generator working group and the ILD software working group for providing the simulation and reconstruction tools and producing the Monte Carlo samples used in this study.
This work has benefited from computing services provided by the ILC Virtual Organization, supported by the national resource providers of the EGI Federation and the Open Science GRID.

\end{document}